\documentclass[prl,aps,twocolumn,superscriptaddress]{revtex4-1}
\usepackage{amsmath, amssymb, amsbsy}
\usepackage{color}
\usepackage{array}
\usepackage{txfonts}
\usepackage{bm, bbm}
\usepackage{graphicx}
\usepackage{appendix}
\usepackage{diagbox, multirow}
\usepackage[percent]{overpic}
\usepackage[colorlinks=true,linkcolor=blue,citecolor=blue,urlcolor=blue,bookmarks=false]{hyperref}
\usepackage[normalem]{ulem}

\begin{document}

\title{Unveiling a critical stripy state in the triangular-lattice SU(4) spin-orbital model}

\author{Hui-Ke Jin}
\affiliation {Department of Physics TQM, Technische Universit\"{a}t M\"{u}nchen, James-Franck-Straße 1, D-85748 Garching, Germany}

\author{Rong-Yang Sun}
\affiliation{Kavli Institute for Theoretical Sciences, University of Chinese Academy of Sciences, Beijing 100190, China}
\affiliation{Computational Materials Science Research Team, RIKEN Center for Computational Science (R-CCS), Kobe, Hyogo, 650-0047, Japan}
\affiliation{Quantum Computational Science Research Team, RIKEN Center for Quantum Computing (RQC), Wako, Saitama, 351-0198, Japan}

\author{Hong-Hao Tu}
\email{hong-hao.tu@tu-dresden.de}
\affiliation{Institut f\"ur Theoretische Physik, Technische Universit\"at Dresden, 01062 Dresden, Germany}

\author{Yi Zhou}
\email{yizhou@iphy.ac.cn}
\affiliation {Institute of Physics, Chinese Academy of Sciences, Beijing 100190, China}
\affiliation{Songshan Lake Materials Laboratory, Dongguan, Guangdong 523808, China}
\affiliation{Kavli Institute for Theoretical Sciences, University of Chinese Academy of Sciences, Beijing 100190, China}
\affiliation{CAS Center for Excellence in Topological Quantum Computation, University of Chinese Academy of Sciences, Beijing 100190, China}
\date{\today}

\begin{abstract}
The simplest spin-orbital model can host a nematic spin-orbital liquid state on the triangular lattice. We provide clear evidence that the ground state of the SU(4) Kugel-Khomskii model on the triangular lattice can be well described by a ``single'' Gutzwiller projected wave function with an emergent parton Fermi surface, despite it exhibits strong finite-size effect in quasi-one-dimensional cylinders. The finite-size effect can be resolved by the fact that the parton Fermi surface consists of open orbits in the reciprocal space. Thereby, a stripy liquid state is expected in the two-dimensional limit, which preserves the SU(4) symmetry while breaks the translational symmetry by doubling the unit cell along one of the lattice vector directions. It is indicative that these stripes are critical and the central charge is $c=3$, in agreement with the SU(4)$_1$ Wess-Zumino-Witten conformal field theory. All these results are consistent with the Lieb-Schultz-Mattis-Oshikawa-Hastings theorem.
\end{abstract}

\maketitle

{\em Introduction.---} 
The search of quantum spin liquids is one of central issues in condensed matter physics~\cite{Savary2016,Zhou2017,Knolle2019,Broholm2020}. 
The quantum fluctuations and geometric frustration are two paramount mechanisms for suppressing magnetic order.
With both mechanisms in play, the $S=1/2$ Heisenberg antiferromagnet on the triangular lattice was regarded as a natural candidate hosting a quantum spin liquid. While Anderson proposed a disordered ground state for this model~\cite{Anderson1973}, such scenario was, however, ruled out by later studies~\cite{Huse1988,Jolicoeur1989}. 
By now its has been established that the ground state of the $S=1/2$ Heisenberg antiferromagnet on the triangular lattice exhibits a three-sublattice $120^{\circ}$ magnetic order~\cite{White2007}. In a sense, $S=1/2$, the most quantum case for SU(2) spins, is still not ``quantum'' enough to melt the classical order on the triangular lattice.

Apart from additional competing interactions~\cite{Block2011,Mishmash2013,Zhu2015,Hu2015,Wietek2017,Gong2017,Saadatmand2017,Hu2019} and stronger geometric frustration~\cite{Yan2011,Depenbrock2012,Liao2017,He2017}, orbital degeneracy provides an alternative mechanism to enhance quantum fluctuations~\cite{Feiner1997}. 
In certain transition metal oxides, spin and orbital degrees of freedom might play a symmetric role and be promoted to an SU(4) symmetry. The SU(4) Kugel-Khomskii (KK) model is a minimal model for describing such spin-orbital materials~\cite{Kugel1982,Li1998,Pati1998,Khaliullin2000,Tokura2020,Reynaud2001,Yamada2018}, where the larger symmetry amplifies quantum fluctuations and might stabilize spin-orbital liquid ground states. The SU($N$) quantum magnetism has also been studied in the platform of ultracold atoms in optical lattices~\cite{Hermele2009,Gorshkov2010}. 
Meanwhile, it was proposed that a Hubbard-type model with an approximate SU(4) symmetry is feasible to depict correlated moir\'{e} materials~\cite{Xu2018,Po2018,Zhang2019,Wu2019,Schrade2019}. 

From theoretical side, much effort has been devoted to investigating quantum phases in SU(4) quantum magnets. In one dimension, the SU(4) KK model is integrable and has gapless excitations~\cite{Sutherland1975,Li1999}, whose low-energy effective theory is the SU(4)$_1$ Wess-Zumino-Witten (WZW) conformal field theory (CFT)~\cite{Affleck1986,Azaria1999,fuhringer2008}. The ground state of the SU(4) KK model on a two-leg ladder breaks the translational symmetry with the formation of SU(4)-singlet plaquettes~\cite{Bossche2001,Weichselbaum2018}. With extra interactions, the SU(4)-singlet plaquette state becomes the exact ground state on the ladder~\cite{Chen2005}. The situation in two dimensions is less well studied: Several complementary methods point to a Dirac-type spin-orbital liquid for the SU(4) KK model on the honeycomb lattice~\cite{Corboz2012,Andrade2019}, whereas a few candidate ground states, such as the plaquette order~\cite{Li1998,Bossche2000}, a $\mathbb{Z}_2$ spin-orbital liquid~\cite{Wang2009}, and simultaneous dimerization with broken SU(4) symmetry~\cite{Corboz2011}, have been proposed for the SU(4) KK model on the square lattice.

The ground state of the SU(4) KK model on the triangular lattice also remains elusive. Exact diagonalization and variational calculations on finite-size clusters indicate an SU(4)-singlet plaquette liquid~\cite{Penc2003}. 
Recently, a density matrix renormalization group (DMRG) study, in combination with field theory analysis, suggested a gapless liquid ground state with an emergent parton Fermi surface~\cite{Xu2020}. 
On the contrary, another DMRG study~\cite{Sheng2021} and a self-consistent mean-field theory~\cite{Chen2021} favor a stripe ordered state.

In this work, we reexamine the ground state of the SU(4) KK model on the triangular lattice by combining the DMRG method~\cite{White1992,White1993} with Gutzwiller projected wave function approach. Our recent methodology development~\cite{Tu2020,Jin2020,Jin2020_2} provides a useful tool for converting Gutzwiller projected wave functions into matrix product states (MPSs) and exploiting them to boost DMRG calculations and characterize ground states { (see similar methods in Refs.~\cite{petrica2020,aghaei2020})}. With two complementary methods and extensive numerical efforts on cylinders with circumference up to $L_y=8$, we have gathered clear evidence for unveiling a critical stripy ground state in the two-dimensional (2D)  limit. This state preserves SU(4) symmetry, but spontaneously breaks translational symmetry by doubling the unit cell along one of the lattice vector directions. 
It is worth emphasizing that there is a marvelous way to approach the 2D limit: although our DMRG results display strong finite size effect in narrow cylinders, we find that all these ground states can be well depicted by a ``single" Gutzwiller projected wave function when the circumference is even. This microscopic wave function possesses a parton Fermi surface that consists of open orbits in the reciprocal space, and its quality is verified by DMRG through variational energy and wave function fidelity. We also observe the even-odd discrepancy in narrow cylinders, as pointed out in Ref.~\cite{Xu2020}. However, with a much larger bond dimension and higher numerical precision, our DMRG calculations reveal that  the ground states on $L_y=3$ cylinders are gapped. This result disagrees with the gapless conclusion made in Ref.~\cite{Xu2020}, while can be naturally interpreted as a result of the non-trivial topology of the parton Fermi surface.

\begin{figure}[tb]
	\centering
	\includegraphics[width=\linewidth]{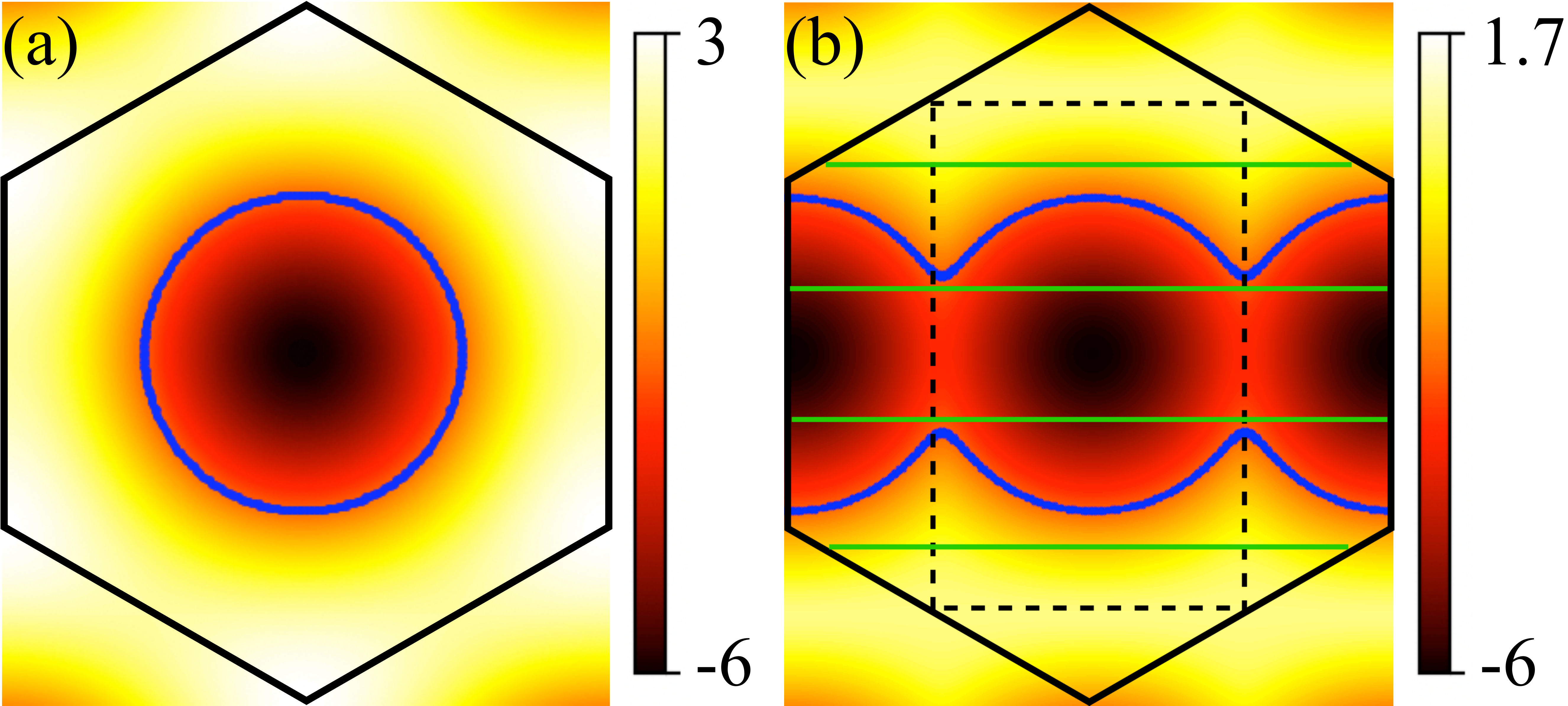}
	\caption{(Color online) (a) The single-particle spectrum as a contour plot for the uniform $\pi$-flux parton state~\eqref{eq:Hpi} in the first Brillouin zone. The blue circle stands for the Fermi surface at 1/4 filling. (b) The single-particle spectrum for the lower band of the stripy parton state~\eqref{eq:Hstripy} with $\delta=0.15$ in the (unfolded) first Brillouin zone. The blue curve stands for the deformed Fermi surface at 1/4 filling, which consists of open orbits in the reciprocal space. Dashed lines enclose the folded Brillouin zone. Green lines represent momenta on XC4 cylinders that are allowed by the APBC along $y$ direction for partons.}
	\label{fig:fermisurface}
\end{figure}

{\em Model Hamiltonian.---}
The SU(4) KK model on the triangular lattice is defined by
\begin{equation}
	H = \frac{1}{2}\sum_{\langle i j \rangle} \left(  {\bm \sigma}_i \cdot {\bm
		\sigma}_j +1 \right) \left(  {\bm \tau}_i \cdot {\bm \tau}_j +
	1 \right),    \label{eq:SU4Model}
\end{equation}
where $\langle{}ij\rangle$ denotes a nearest-neighbor (NN) bond and $\bm{\sigma}$ ($\bm{\tau}$) represents Pauli matrices for spin (orbital) degrees of freedom.
The four-dimensional Hilbert space at each site forms the fundamental representation of the SU(4) Lie algebra, and $\{\sigma^a,\tau^b,\sigma^a\tau^b\}$ are linear combinations of the fifteen SU(4) generators $\bm{\lambda}\equiv(\lambda^1,\lambda^2,\cdots,\lambda^{15})$ normalized with $\mathrm{tr}(\lambda^\mu \lambda^\nu)=2\delta_{\mu\nu}$~\cite{Georgi}, with which the Hamiltonian \eqref{eq:SU4Model} can be rewritten in an explicitly SU(4)-invariant form $H=\sum_{\langle{}ij\rangle}(\bm{\lambda}_{i}\cdot\bm{\lambda}_{j} +1/2)$.

{\em Fermionic parton wave function.---}
For SU($N$) quantum magnets, parton constructions have often been employed to derive effective theories and construct trial wave functions~\cite{Wang2009,Corboz2012,Xu2020,Yamada2021}.
We follow this strategy by introducing the SU(4) fermionic parton representation,
\begin{align}
	\sigma^a_{i} \rightarrow \bm{f}^\dagger_i\sigma^a\tau^0\bm{f}_i,~\tau^b_{i}\rightarrow \bm{f}^\dagger_i\sigma^0\tau^b\bm{f}_i,~(\sigma^a\tau^b)_i \rightarrow \bm{f}^\dagger_i\sigma^a\tau^b\bm{f}_i,
\end{align}
where $\sigma^0$ ($\tau^0$) is the identity matrix in spin (orbital) space, and $\bm{f}_i=(f_{i,1},\cdots,f_{i,4})^T$ are fermion annihilation operators ($\bm{f}^\dagger_i$: corresponding creation operators). In order to restore the four-dimensional physical Hilbert space, a local single-occupancy constraint, $\sum_{m=1}^4 f^\dagger_{i,m}f_{i,m}=1$, has to be imposed.

\begin{figure}
	\includegraphics[width=\linewidth]{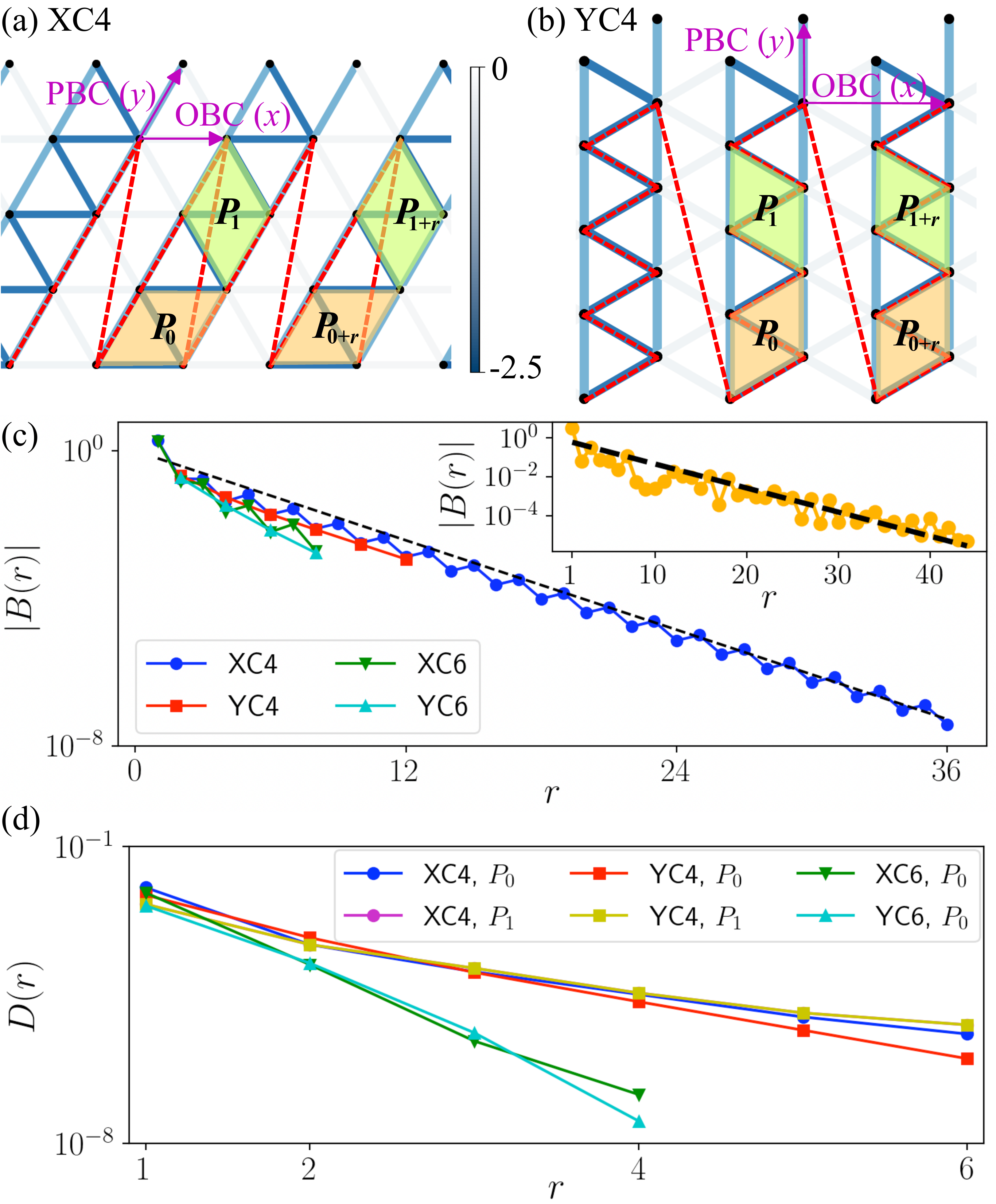}
	\caption{(Color online) The mapping of the triangular lattice to a 1D chain on (a) XC and (b) YC cylinders. The colors of the bonds represent the expectation value $\Lambda_{ij}\equiv\langle\bm{\lambda}_i\cdot\bm{\lambda}_j\rangle$ between neighboring sites for the central columns in a cylinder of circumference $L_y=4$ and length up to $L_x=24$. The red dashed lines represent the 1D path in DMRG calculations, and plaquette SU(4)-singlet projectors are defined on the rhombuses.
	(c) The SU(4) spin correlation function $B(r)=\Lambda_{i,i+r\hat{x}}$ versus distance $r$ for XC and YC cylinders with length $L_x=48$. The dashed line indicating an exponential decay of $|B(r)|$ is a guide to the eye. Inset: $B(r)$ versus $r$ for an XC3 cylinder with length $L_x=60$. The dashed line indicates an exponential decay of $|B(r)|$.
	(d)	The two-plaquette SU(4)-singlet correlation function $D(r)\equiv{}\langle{}{P}_{p}{P}_{p+2r\hat{x}}\rangle-\langle{}{P}_{p}\rangle\langle{}{P}_{p+2r\hat{x}}\rangle$ versus distance $r$ in cylinders of length $L_x=24$ (XC4 and YC4) and of length $L_x=16$ (XC6 and YC6).
	}
	\label{fig:fig1}
\end{figure}

With the help of the parton representation, a mean-field decomposition can be carried out to obtain various quadratic forms of partons, dubbed ``effective Hamiltonian", which determines mean-field ground states and low-energy excitations. As an example, the uniform $\pi$-flux state, suggested in Ref.~\cite{Xu2020}, is described by the following effective Hamiltonian: 
\begin{equation}
	H_{\rm \triangle=\pi{}} = -\sum_{m=1}^4\sum_{\langle ij \rangle} \left( f^\dag_{i,m}f_{j,m} + f^\dag_{j,m}f_{i,m}\right). \label{eq:Hpi}
\end{equation}
This parton Hamiltonian~\eqref{eq:Hpi} preserves SU(4) rotational and lattice translational symmetries, and gives rise to a fourfold degenerate band structure.
The single-occupancy or 1/4 filling is indicative of a circular Fermi surface in the parton band [see Fig.~\ref{fig:fermisurface}(a)] and a gapless state.
In this work, we propose a stripy state as a result of the instability of this uniform $\pi$-flux state. This stripy state breaks the lattice translational symmetry by doubling the unit cell along the one of the lattice vector directions. Without loss of generality, we take $x$ direction to be the symmetry breaking direction [see Figs.~\ref{fig:fig1}(a) and \ref{fig:fig1}(b)]. The mean-field Hamiltonian for such a stripy state reads
\begin{equation}
	H_{\rm stripy}(\delta) = -\sum_{m=1}^4\sum_{\langle ij \rangle} \left[1+(-1)^{r_i}\delta\right]\left( f^\dag_{i,m}f_{j,m} + f^\dag_{j,m}f_{i,m} \right), \label{eq:Hstripy}
\end{equation}
where $r_i$ is the column index of lattice site $i$, $r_{j}\ge{}r_{i}$ is assumed for NN bond $\langle{}ij\rangle$, and $\delta{}\in[-1,1]$ is the stripy strength.
A finite $\delta$ deforms the original circular Fermi surface, which recovers the uniform $\pi$-flux state at $\delta{}=0$. When $\delta>\delta_{c}\approx0.09$, the deformed Fermi surface consists of open orbits in the reciprocal space [see Fig.~\ref{fig:fermisurface}(b)].

Thanks to the newly developed method~\cite{Tu2020,Jin2020}, the ground states of Hamiltonians~\eqref{eq:Hpi} and \eqref{eq:Hstripy} can be represented as MPSs, upon which the Gutzwiller projection can be easily implemented to impose the single-occupancy constraint and obtain many-body wave functions in the physical spin-orbital Hilbert space. These trial wave functions, in their MPS form, can be used to initialize DMRG calculations, which is dubbed as ``Gutzwiller-boosted DMRG''~\cite{Jin2020_2}. Consequently, both DMRG-obtained ground states and Gutzwiller projected trial wave functions can be flexibly studied with powerful MPS techniques. 

To perform DMRG calculations on triangular lattices, we consider cylinder geometries, for which the periodic boundary condition (PBC) for spin-orbital degrees of freedom is imposed along the $y$ direction, while the $x$ direction is left open.
Following the common practice of DMRG, the 2D model is mapped to a one-dimensional (1D) model with long-range interactions. In choosing the 1D path, we consider two different mappings, labeled as XC$L_y$ and YC$L_y$, where $L_y$ is the circumference of the cylinder [see Figs.~\ref{fig:fig1}(a) and \ref{fig:fig1}(b)].

During our numerical calculations, we have made explicit use of the U(1)$^{\otimes3}$ quantum numbers corresponding to the Cartan generators $\{\sigma^z,\tau^z,\sigma^z\tau^z\}$ (see the Supplemental Material for more details).
Our DMRG calculations have been performed on cylinders with circumference up to $L_y=6$.
The number of states kept in DMRG calculations is as large as $D=18000$, resulting in a truncation error $\epsilon_{\text{D}}\simeq10^{-6}$ ($\epsilon_{\text{D}}\simeq10^{-4}$) for $L_y=4$ ($L_y=5, 6$, and $8$).
In the processes of converting Gutzwiller projected wave functions to MPS~\cite{Tu2020,Jin2020,Jin2020_2}, the bond dimension is kept up to $\tilde{D}=16000$, with an accumulated truncation error $\epsilon_{\rm M}<0.05$.
It is noted that we can choose antiperiodic boundary condition (APBC) for partons along the compactified $y$ direction in addition to PBC, which still yields PBC for spin-orbital degrees of freedom.

\begin{figure}
	\centering
	\includegraphics[width=\linewidth]{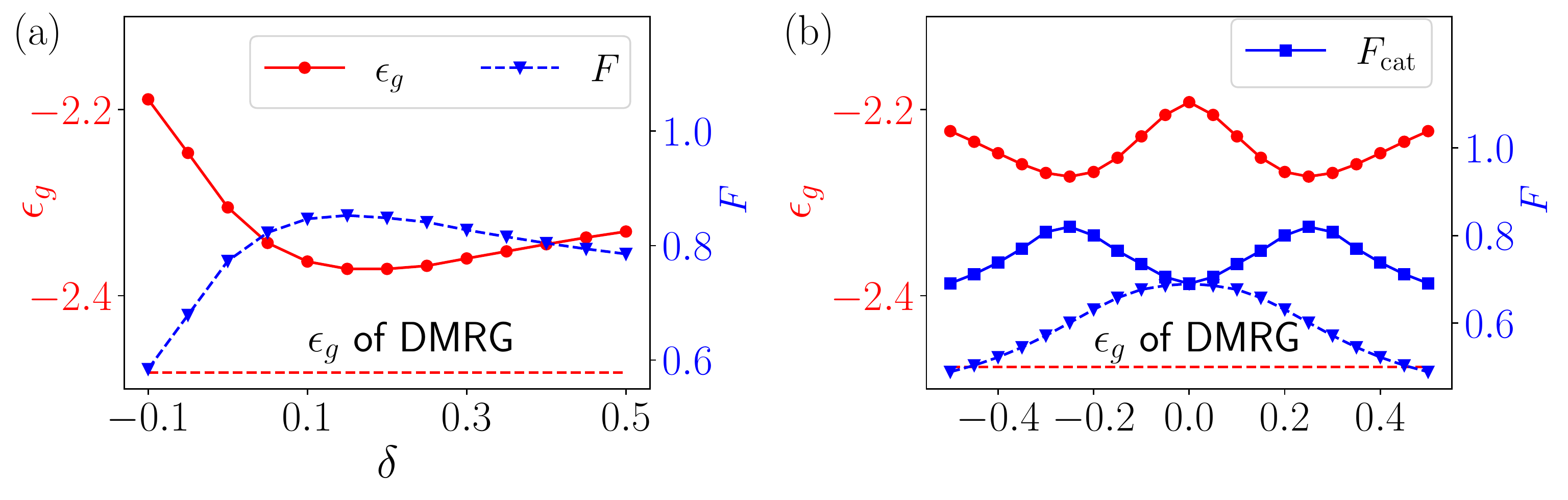}
	\caption{(Color online) The variational energy per site $\epsilon_g(\delta)$ (left y-axis) of the Gutzwiller projected state $\vert\Psi_{\mathrm{stripy}}(\delta)\rangle$ and the wave-function fidelity $F(\delta)$ (right y-axis) versus $\delta$ for (a) an XC4 cylinder with $L_x=8$ and (b) an XC4 torus with $L_x=6$. For the torus case, the fidelity $F_\mathrm{cat}$ is computed with respect to one of the cat states $|\Psi_{\rm cat}(\delta)\rangle \propto  \vert\Psi_{\mathrm{stripy}}(\delta)\rangle \pm \vert\Psi_{\mathrm{stripy}}(-\delta)\rangle$ having larger overlap with the DMRG-obtained wave function $\vert\Psi_\mathrm{DMRG}\rangle$. The MPS bond dimensions for representing the Gutzwiller projected states are (a) $\tilde{D}=6000$ and (b) $\tilde{D}=16000$, ensuring a small accumulated truncation error $\epsilon_\mathrm{M}<0.025$. The reference energy $\epsilon_g$ (red dashed lines) and the wave function $\vert\Psi_\mathrm{DMRG}\rangle$ for computing the fidelity $F(\delta)$ are obtained using Gutzwiller-boosted DMRG with $D=15000$ for both (a) and (b). 
	}\label{fig:fig2}
\end{figure}

{\em Energetics and fidelity of the stripy state.---} By the combinative method of DMRG and Gutzwiller projected wave functions, we have obtained the ground states of the KK model~\eqref{eq:SU4Model} with different cylindrical boundary conditions and various system sizes.
As mentioned earlier, the Gutzwiller projected wave function $|\Psi_{\rm stripy}(\delta{})\rangle$ is determined from the effective Hamiltonian~\eqref{eq:Hstripy} at 1/4 filling. As illustrated in Fig.~\ref{fig:fig2}(a), its variational energy per site $\epsilon_g(\delta)$ (on the cylinder) reaches its minimum at $\delta{}\approx0.15$, suggesting that the uniform $\pi$-flux state ($\delta=0$) is unstable against the stripy deformation. 
The wave-function fidelity between $|\Psi_{\rm stripy}(\delta{})\rangle$ and the DMRG-obtained ground state $|\Psi_{{\rm DMRG}}\rangle$, $F(\delta{})\equiv\vert\langle\Psi_{\rm stripy}(\delta{})|\Psi_{{\rm DMRG}}\rangle\vert$, is also maximized at $\delta{}\approx0.15$, in good agreement with the energy comparison. 
Here $|\Psi_{\rm DMRG}\rangle$ is obtained by using Gutzwiller-boosted DMRG initialized with $|\Psi_{\rm stripy}(\delta{}=0.15)\rangle$.

As is well known, boundary effects due to the cylinder geometry might lead to biased results.
To eliminate boundary effects, we also consider the Gutzwiller projected wave function $|\Psi_{\rm stripy}(\delta{})\rangle$ on an XC torus with $L_{x}\times{}L_{y}=6\times{}4$.
It is found that a lower variational energy is achieved when partons assume APBCs along both directions. The numerical results for the variational energy $\epsilon_g(\delta)$ and the fidelity $F(\delta{})$ are shown in Fig.~\ref{fig:fig2}(b). Note here the reference $|\Psi_{\rm DMRG}\rangle$ is also obtained by using Gutzwiller-boosted DMRG initialized with $|\Psi_{\rm stripy}(\delta{}=0.25)\rangle$. 
At first glance, there is a contradiction between $\epsilon_g(\delta)$ and $F(\delta)$: the lowest energy $\epsilon_g(\delta)$ is achieved at $\delta\approx\pm{}0.25$, while the highest fidelity $F(\delta)$ appears at $\delta=0$. However, this seemingly contradiction can be resolved as follows: On a torus, $|\Psi_{\rm stripy}(\delta)\rangle$ and its dual stripy state $|\Psi_{\rm stripy}(-\delta)\rangle$ are energetically degenerate. For $|\delta|\ge{}0.25$, these two states are found to be almost orthogonal to each other, with an overlap less than $0.05$. 
Thus, one can construct cat states, $|\Psi_{\rm cat}(\delta)\rangle \propto |\Psi_{\rm stripy}(\delta)\rangle\pm|\Psi_{\rm stripy}(-\delta)\rangle$ and compute the fidelity $F_{\rm cat}(\delta)\equiv\vert\langle{}\Psi_{\rm cat}(\delta)|\Psi_{\rm DMRG}\rangle\vert$ (defined by the larger overlap with one of the two cat states). As shown in Fig.~\ref{fig:fig2}(b), the fidelity $F_{\rm cat}(\delta)$ reaches its maximum at $\delta\approx\pm{}0.25$, which is indeed consistent with the behavior of $\epsilon_g(\delta)$.
We emphasize that the same results in Fig.~\ref{fig:fig2} has also be obtained by using randomly initialized DMRG calculations.

It is remarkable that the unit-cell doubling in the stripy state does not open an energy gap on the parton Fermi surface. This remaining parton Fermi surface is due to the 1/4 filling of the Fermi sea, and is quite different from the usual gap opening by the dimerization in SU(2) quantum spin-1/2 systems, on which the half-filled spinon band acquires an energy gap by the dimerization.

{\em Correlation functions.---}
To characterize the stripy state, we have calculated a variety of correlation functions. The leading-order correlation consists of two SU(4) spins, i.e., the SU(4) spin correlation function $\Lambda_{ij}\equiv\langle\bm{\lambda}_i\cdot\bm{\lambda}_j\rangle$, whose values on NN bonds, $\Lambda_{\langle{}ij\rangle}\in[-2.5,1.5]$, give rise to the local energy density up to a constant of 1/2.
It can be seen from Figs.~\ref{fig:fig1}(a) and \ref{fig:fig1}(b) that, on both types of cylinders, XC4 and YC4, $\Lambda_{\langle{}ij\rangle}$ exhibits sizable twofold oscillations along the $x$ direction, indicating translational symmetry breaking. (Strictly speaking, for the YC4 cylinder, the ground state is still translationally invariant, as the YC4 cylinder itself has a doubly enlarged unit cell.)
Meanwhile, the translational symmetry is preserved along the periodic $y$ direction, which allows us to define a $y$-independent correlation function, $B(r)=\Lambda_{i,i+r\hat{x}}$. Numerical results from DMRG suggest an exponential decay of $|B(r)|$ on cylinders up to $L_{x}=48$, as demonstrated in Fig.~\ref{fig:fig1}(c), which is consistent with the scenario of a parton Fermi surface consisting of open orbits in the reciprocal space. This is due to the fact that none of the allowed momenta in these finite $L_y$ cylinders cuts  the Fermi surface [see Fig.~\ref{fig:fermisurface}(b)]. Note that the APBC (along $y$ direction) for partons leads to lower energy states on $L_y=4$ cylinders, while the PBC is energetically favored by $L_y=6$ cylinders.
We have also performed DMRG calculations on odd circumference ($L_y=3$) cylinders (see the Supplemental Material for more details). The correlation function $|B(r)|$ also exhibits an exponential decay on a $L_y=3$ cylinder, whose correlation length is about twice larger than those on the $L_y=4$ and 6 cylinders. Note that this is inconsistent with the conclusion drawn in Ref.~\cite{Xu2020}, but agrees to the scenario of a open parton Fermi.

In addition to the two-site SU(4) spin correlation function $\Lambda_{ij}$, two-plaquette SU(4)-singlet correlation functions have been investigated as well. To form an SU(4)-singlet, at least four SU(4) spins are required. In particular, we are interested in four-site plaquettes on strong double-columns, which have stronger bonds or lower energy density $\Lambda_{\langle{}ij\rangle}$. Thus, the two-plaquette SU(4)-singlet correlation function can be defined on strong double-columns as $D(r)\equiv\langle{}P_p{}P_{p+2r\hat{x}}\rangle-\langle{}P_p{}\rangle\langle{}P_{p+2r\hat{x}}\rangle$, where $p$ labels a four-site plaquette on strong double-columns and the operator $P_p$ projects a state in the plaquette-$p$ onto the SU(4) singlet.
The stripy structure breaks the $C_3$ lattice rotational symmetry and the cylindrical geometry breaks the mirror symmetry, resulting in two distinct types of plaquettes on a single XC or YC cylinder, i.e., $P_0$ and $P_1$ illustrated in Figs.~\ref{fig:fig1}(a) and \ref{fig:fig1}(b), respectively.
The expectation values of SU(4)-singlet projectors on strong double-columns (in the bulk) are given by $\langle{}P_0\rangle\approx\langle{}P_1\rangle\approx0.32~(0.29)$ on XC4 (XC6) and YC4 (YC6) cylinders.  
The two-plaquette correlation function $D(r)$ is shown in Fig.~\ref{fig:fig1}(d), which exhibits a rapid decay as the distance $r$ increases.

\begin{figure}
	\centering
	\includegraphics[width=\linewidth]{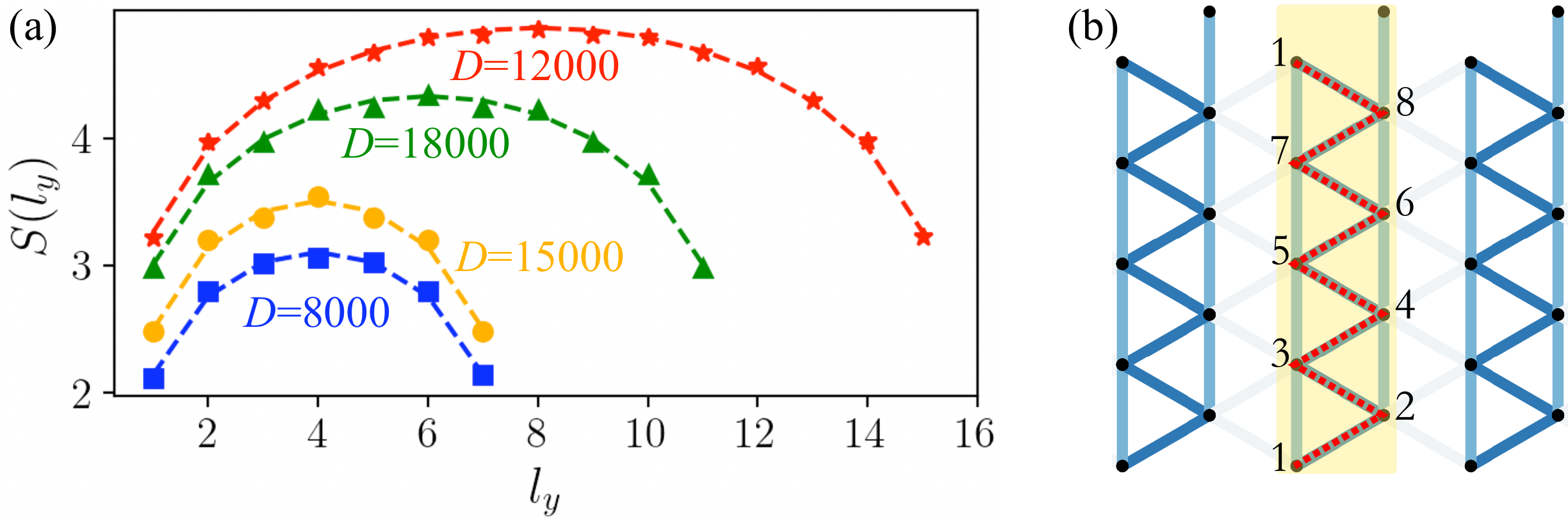}
	\caption{(Color online)  (a) Entanglement entropy $S(l_y)$ versus subsystem length $l_y$ on a strong double-column of $2L_y$ sites at the center of cylinders with length $L_x=16$. The blue squares (orange dots) stand for the results obtained from $|\Psi_{\rm stripy}(\delta=0.15)\rangle$ ($|\Psi_{\rm DMRG}\rangle$) on a YC4 cylinder. The green triangles stand for the results from $|\Psi_{\rm DMRG}\rangle$ on a YC6 cylinder and the red stars for those from $|\Psi_{\rm DMRG}\rangle$ on a YC8 cylinder. The fittings of central charges and corresponding standard deviation errors are: 3.0$\pm$0.07 (blue squares), 3.1$\pm$0.10 (orange dots), 2.9$\pm$0.06 (green triangles), and 3.0$\pm$0.06 (red stars). Notice that for a	 better presentation the data of $S(l_y)$ for the red stars have been globally shifted by 0.5. (b) Two columns of eight sites in the YC4 cylinder as a 1D zigzag chain. 
	}\label{fig:columnEE}
\end{figure}

{\em Entanglement entropy.---}
The gapless nature of the stripy state and the emergency of a parton Fermi surface can be further revealed and characterized by the von Neumann entanglement entropy (EE). Intuitively, the stripy state can be thought of as $L_x/2$ weakly coupled 1D chains, i.e., zigzag SU(4) spin chains on the aforementioned strong double-columns [see Fig.~\ref{fig:columnEE}(b)]. We shall focus on YC cylinders [see the mapping of the YC cylinder in Fig.~\ref{fig:fig2}(b)], on which each zigzag chain consists of $2L_y$ sites. 

Within each strong double-column, the entanglement between one subsystem with $l_y$ spins and the remaining $2L_y-l_y$ spins is measured by the von Neumann EE $S(l_y)$. (For instance, for the $i^{th}$ strong double-column, we calculate the von Neumann EE between the $(2iL_y+l_y)^{th}$ and the $(2iL_y+l_y+1)^{th}$ sites, with $l_y=1,\ 2,\ ...,\ 2L_y-1$.) For 1D critical systems described by CFT, $S(l_y)$ follows the scaling law~\cite{Holzhey1994,Vidal2003,Cardy2009}
\begin{equation}
	S(l_y)=\frac{c}{3}\log\left(\frac{2L_y}{\pi}\sin\frac{\pi{}l_y}{2L_y}\right) +\gamma, \label{eq:CFTEE}
\end{equation}
where $c$ is the central charge of the CFT and $\gamma$ is a nonuniversal constant.
We have evaluated the EE $S(l_y)$ for the Gutzwiller projected state $|\Psi_{\rm stripy}(\delta=0.15)\rangle$ as well as the DMRG-obtained state $|\Psi_{\rm DMRG}\rangle$ on a strong double-column in the middle of YC cylinders.  
As shown in Fig.~\ref{fig:columnEE}(a), the EE of both states are well fitted by Eq.~\eqref{eq:CFTEE}, and the extracted central charge is $c\simeq{}3$. Although many CFTs have this central charge, the SU(4)$_1$ WZW model seems to be the most natural candidate, since it is compatible with the SU(4) symmetry and often shows up in SU(4) critical chains. The SU(4)$_1$ criticality is also consistent with the description of 1D four-component gapless fermions subject to Gutzwiller projection~\cite{Kawakami1992}. Although not shown, we have also obtained same results for other zigzag chains in the bulk of YC cylinders.

Treating a cylinder as an effective 1D chain with a big unit cell of $L_y$ spins and dividing the cylinder into two parts, the first $l_x$ big unit cells and the remaining $L_x-l_x$ ones, we are able to calculate the EE $S(l_x)$ between these two subsystems. It is found that as $l_x$ increases, $S(l_x)$ quickly saturates with twofold oscillations on even circumference cylinders (see the Supplemental Material for more details), which is consistent with the short-range correlations along the $x$ direction and the translational symmetry breaking.
 For the $L_y=3$ cylinders, we also find that $S(l_x)$ quickly saturates, however, with fourfold oscillations (see the Supplemental Material for more details).
Though this fourfold oscillation is not in agreement with the unit-cell doubling in our stripy ansatz, we argue that the ground state on a $L_y=3$ system is still more likely a stripy state but with a larger unit cell.
Therefore, all these results on EE point to an emergent parton Fermi surface consisting of open orbits in the SU(4) KK model on the triangular lattice, since a circular Fermi surface must lead to a nonzero central charge of $S(l_x)$.
Note that here $S(l_x)$ changes little with bond dimension $D$ up to $D=18000$, indicating that this saturation of $S(l_x)$ is not caused by truncation errors.

{\em Lieb-Schultz-Mattis-Oshikawa-Hastings argument.---}
The scenario of a critical stripy ground state is also consistent with the Lieb-Schultz-Mattis-Oshikawa-Hastings (LSMOH) theorem~\cite{LSM,LSM2,Oshikawa2000,Hastings2004}: The SU(4) KK model on the triangular lattice, with a fundamental representation $\mathbf{4}$ on each site, cannot have a nondegenerate gapped ground state preserving both SU(4) and translational symmetries. When considering the Young tableaux representing SU(4) representations within each unit cell, such trivial gapped state can only be achieved with the number of boxes in the Young tableaux being an integer multiple of 4. With the stripy deformation in the ground state, the unit cell is doubled and has a Young tableaux of two boxes (symmetric representation $\mathbf{10}$ or antisymmetric representation $\mathbf{6}$). According to the LSMOH argument, this stripy state would either be gapless or have topological order, ruling out the possibility of a trivial gapped state. As our numerical results provide strong evidence of criticality on each strong double-column, a gapped state with topological order is unlikely, which leaves the critical stripy state as the only possibility.

{\em Conclusion and discussion.---}
In summary, we have studied the SU(4) KK model on the triangular lattice by combining the DMRG method with Gutzwiller projected wave functions. With extensive numerical results and analytical arguments, we gather strong evidence of a critical stripy ground state, which spontaneously breaks translational symmetry along one of the lattice vector directions and has an emergent parton Fermi surface. Our proposal of a Gutzwiller projected wave function has been shown to describe well this critical stripy state. We expect that this nematic spin-orbital liquid state is stable in the 2D limit and serves as an excellent ground-state candidate for the SU(4) KK model on the triangular lattice beyond the quasi-1D geometry.
For future works, it would be interesting to understand different instabilities of the parton Fermi surface in the SU(4) KK model. For instance, a chiral spin liquid, which emerges in the presence of three-body interactions~\cite{Lauchli2016,Sheng2021}, is one of such instabilities. This would require to map out the phase diagram near the SU(4) KK model, where other exotic phases of matter might appear. Along this line of research, it would certainly be important to identify suitable experimental realizations of such SU(4) quantum magnets.

{\em Acknowledgment.---}
We thank Zheng Zhu for helpful discussions. Y. Z. is supported by National Natural Science Foundation of China (No. 12034004 and No. 11774306), the K. C. Wong Education Foundation (Grant No. GJTD-2020-01), and the Strategic Priority Research Program of Chinese Academy of Sciences (No. XDB28000000).
H.-K. J. is funded by the European Research Council (ERC) under the European Unions Horizon 2020 research and innovation program (grant agreement No. 771537). H.-H. T. is supported by the Deutsche Forschungsgemeinschaft through project A06 of SFB 1143 (project-id 247310070). The numerical simulations in this work are based on the GraceQ project (\href{gracequantum.org}{www.gracequantum.org}).

{\em Author contributions.---}
The project was conceived by Hong-Hao Tu and Yi Zhou; The numerical calculations were done by Hui-Ke Jin with the help of Rong-Yang Sun; Rong-Yang Sun established and maintained the numeric library in GraceQ project (\href{gracequantum.org}{www.gracequantum.org});  All the authors discussed and analyzed the results; Hui-Ke Jin, Hong-Hao Tu and Yi Zhou wrote the manuscript with input from Rong-Yang Sun.

\bibliography{SU4KKTL}

\newpage

\begin{widetext}
	\setcounter{equation}{0}
	\setcounter{figure}{0}
	\setcounter{table}{0}
	\renewcommand{\theequation}{S\arabic{equation}}
	\renewcommand{\thefigure}{S\arabic{figure}}
	\renewcommand{\thetable}{S\Roman{table}}
	%%%%%%%%%% Prefix a "S" to all equations, figures, tables and reset the counter %%%%%%%%%%
	\newpage
	\begin{center}
		{\bf Supplemental Material for ``Unveiling a critical stripy state in the triangular-lattice SU(4) spin-orbital model''}
	\end{center}
	
	In this Supplemental Material, we provide further details about (1) the convergence diagnosis for DMRG simulations, (2) the efficiency of Gutzwiller-boosted DMRG, (3) the properties of DMRG-obtained ground states, and (4) the boundary effect in the projected uniform $\pi$-flux state.
	
	\section{Convergence diagnosis for DMRG simulations}
	To verify the convergence of DMRG-obtained ground states, we study the scaling behavior of the ground-state energy per site $\epsilon_g$ with the DMRG truncation error $\epsilon_{D}$ on $L_{y} = 6$ cylinders and $L_{y}=8$ cylinder. 
	As shown in Fig.~\ref{fig:sm-dmrg-conv}, $\epsilon_g$ scale linearly with $\epsilon_{\rm D}$ when bond dimension $D \geq 12000$, suggesting a faithful convergence in our DMRG simulations initialized with random MPSs.
	Notice that here we have also shown the results for the Alternative XC (AXC) cylindrical geometry. The details of AXC cylinders will be discussed below.
	\begin{figure}[ht]
		\centering
		\includegraphics[width=0.8\linewidth]{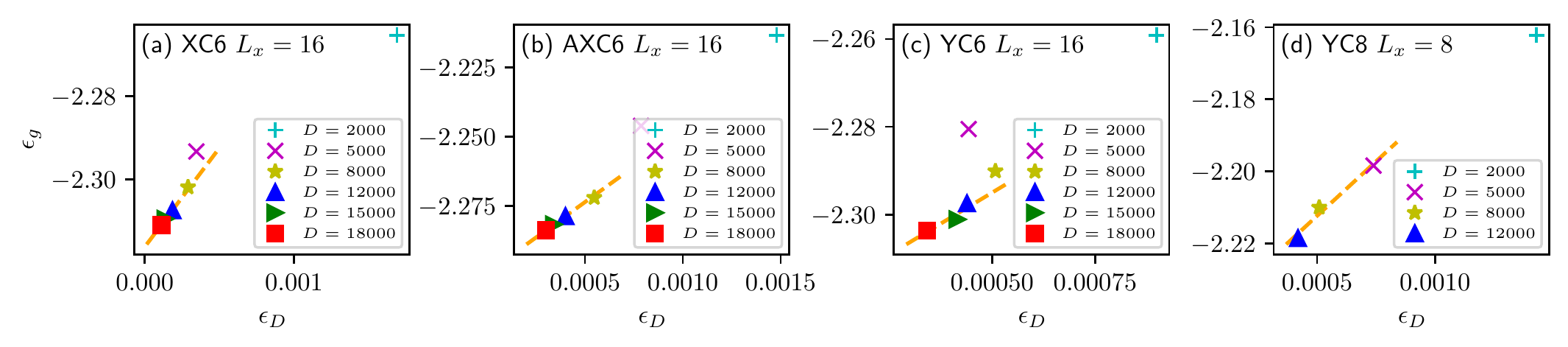}
		\caption{\label{fig:sm-dmrg-conv} Ground-state energy per site $\epsilon_g$ as a function of DMRG truncation error $\epsilon_{\rm D}$ for (a) XC6 with $L_{x} = 16$, (b) AXC6 with $L_{x} = 16$, (c) YC6 with $L_{x} = 16$, and (d) YC8 with $L_{x}=16$. The orange dash lines being a guide to the eye are obtained by a linear fit of the last three data points.}
	\end{figure}

	\section{Efficiency of the Gutzwiller-boosted DMRG}
	\begin{figure}[ht]
		\centering
		\includegraphics[width=0.5\linewidth]{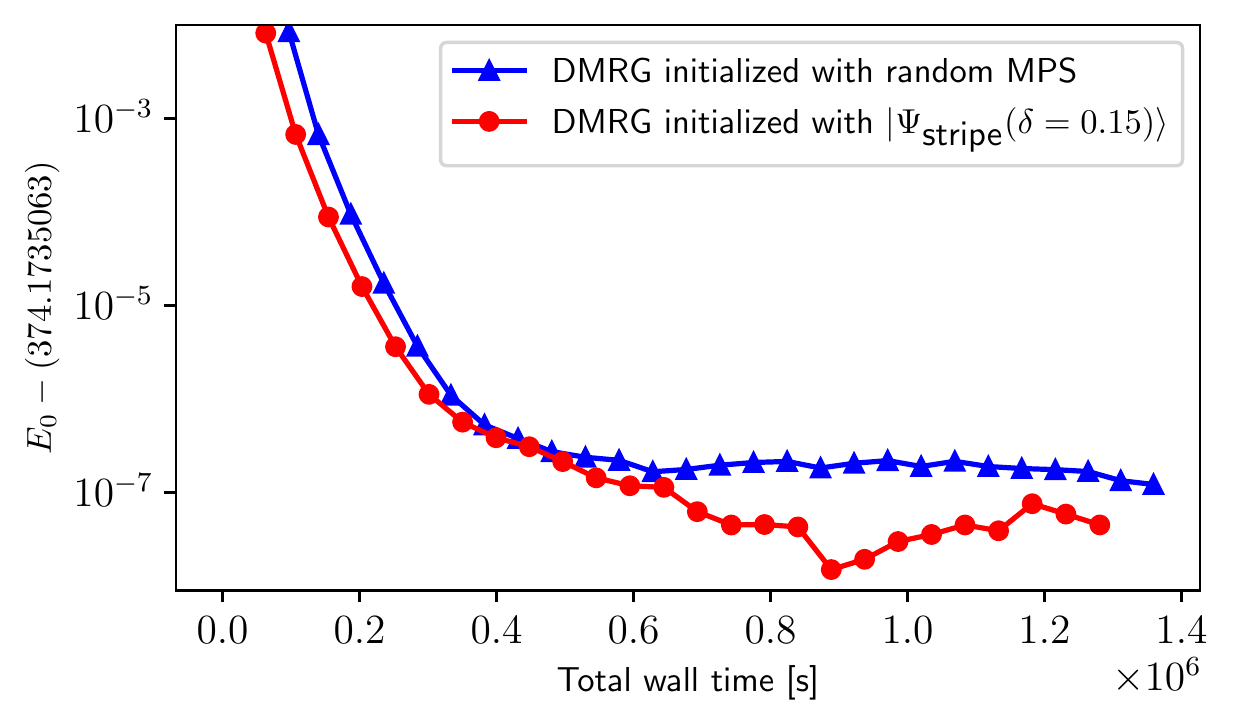}
		\caption{\label{fig:sm-eff-gb-drmg} The ground-state energy (minus a constant) for a specific DMRG sweep with the total wall time cost after the sweep. 
			Here $|\Psi_{\rm{}stripy}(\delta=0.15)\rangle$ has been converted to an MPS with bond dimension $\tilde{D} = 2000$. 
			The time cost of preparing such an initial MPS of  $|\Psi_{\rm{}stripy}(\delta=0.15)\rangle$ can be negligible comparing to the time cost of DMRG calculations.
			Red dots and blue triangles show the energy decreasing of the DMRG simulations initialized with $|\Psi_{\rm stripy}(\delta = 0.15)\rangle$ at $\tilde{D} = 2000$ and a randomly generated MPS, respectively. 
			The lattice geometry is XC4 with $L_{x} = 24$ and the truncation error is fixed to $\epsilon_{\rm D} = 1.0\times 10^{-6}$.}
	\end{figure}
	In order to investigate the quality of the Gutzwiller projected wave function generated from $H_{\rm stripy}(\delta)$ [see Eq.~(4) in the main text], we compare the performance of DMRG simulations with initial MPSs prepared from i) a randomly generated product state and ii) a Gutzwiller projected states $|\Psi_{\rm stripy}(\delta=0.15)\rangle$, respectively.
	In Fig.~\ref{fig:sm-eff-gb-drmg}, we show the optimized ground-state energies as a function of the total wall time for these two DMRG calculations with different initial ansatz.
	At any specific time during DMRG simulations, the MPS obtained by Gutzwiller-boosted DMRG always has a lower energy than that obtained by DMRG initialized with a random MPS. 
	Moreover, a smoother energy descent curve of Gutzwiller-boosted DMRG also indicates a better variational optimization path in the Hilbert space. 
	The wave-function fidelity between the states obtained by these two kinds of DMRG simulations is about $F\approx{}1-10^{-6}$.

	\section{Ground-state properties of the SU(4) Kugel-Khomskii model on the triangular lattice}
	
	\subsection{Entanglement entropy scaling along the $x$ direction}
	By viewing a $L_x\times{}L_y$ cylinder as an effective 1D chain with a unit cell of $L_y$ sites, we have calculated the bipartite entanglement entropy (EE) $S(l_{x})$ for this effective 1D chain with $L_x$ unit cells. 
	The bipartite EE $S(l_x)$ versus the subsystem length $l_{x}$ for different circumferences $L_y$ are shown in Fig.~\ref{fig:sm-eex}.
	We find that as $l_x$ increases, $S(l_x)$ quickly saturate (in the presence of oscillations), independent of cylindrical geometries, suggesting short-range correlations along the $x$ direction in which the translational symmetry is broken.
	Besides, the twofold oscillation for cylinders with $L_y = 4$ [Fig.~\ref{fig:sm-eex}(b)], $L_y = 6$ [Fig.~\ref{fig:sm-eex}(c)], and $L_y=8$ [Fig.~\ref{fig:sm-eex}(d)], and the fourfold oscillation for cylinders with $L_y = 3$ [Fig.~\ref{fig:sm-eex}(a)] indicate the translational symmetry breaking in the $x$ direction.
	\begin{figure}[ht]
		\centering
		\includegraphics[width=0.8\linewidth]{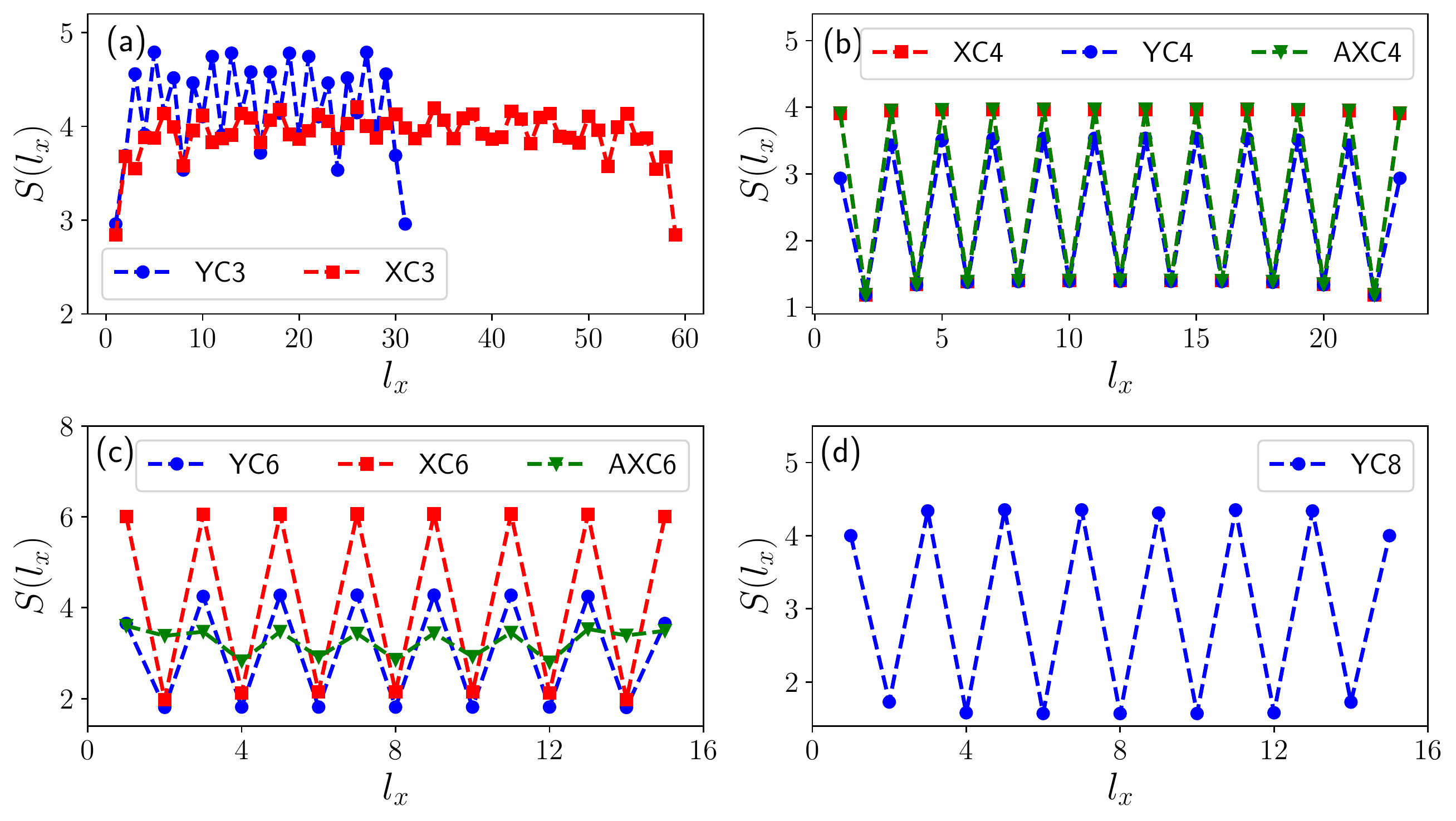}
		\caption{\label{fig:sm-eex} The bipartite EE $S(l_{x})$ versus subsystem length $l_{x}$ on the $L_{y} =$ (a) 3, (b) 4, (c) 6, and (d) 8 cylinders. The type of the cylinders is shown in the legends.
		}
	\end{figure}
	
	\subsection{Alternative XC geometry}
	In addition to the XC and YC cylinders [see Figs.~2(a) and (b) in the main text], we have also considered a 1D chain mapping labeled as AXC$L_y$ [see Fig.~\ref{fig:sm-alter-xc}(a)] for carrying out DMRG simulations.
	We find that $\Lambda_{\langle{}ij\rangle}$, the expectation value of $\bm{\lambda}_i\cdot\bm{\lambda}_j$ on the nearest-neighbor bond $\langle{}ij\rangle$, exhibit an almost uniform pattern in the bulk of an AXC4 cylinder [see Fig.~\ref{fig:sm-alter-xc}(a)], but we still have evidence for a stripy state on AXC cylinders: i) The EE $S(l_x)$ for the AXC geometry also exhibits sizable twofold oscillations, as shown in Figs.~\ref{fig:sm-eex}(b) and (d), indicating a translational symmetry breaking in the $x$ direction; ii) The plaquette-plaquette correlation $D(r)$ also shows a twofold oscillation [see Fig.~\ref{fig:sm-alter-xc}(b)].
	\begin{figure}[ht]
		\centering
		\includegraphics[width=0.8\linewidth]{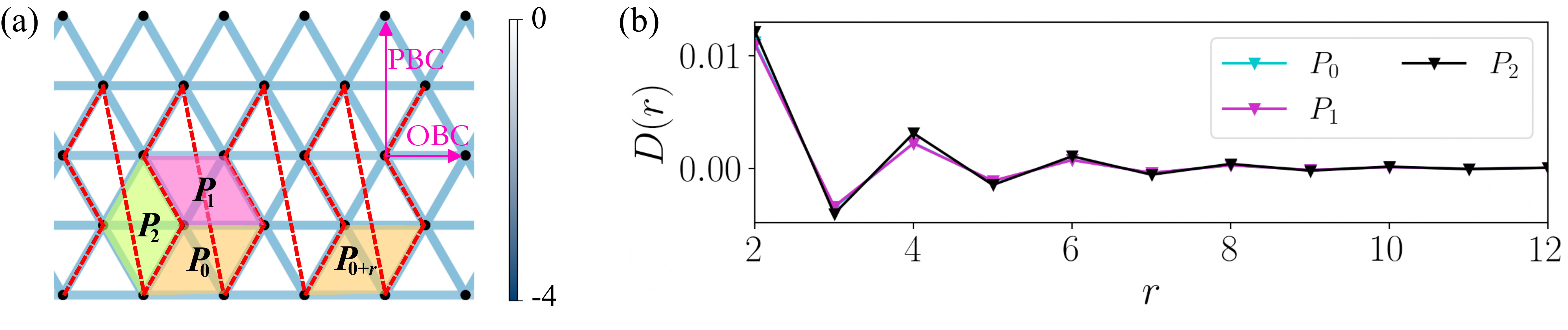}
		\caption{\label{fig:sm-alter-xc} (a) The mapping of the triangular lattice to a 1D chain on an AXC cylinder. The colors of the bonds represent the expectation value $\Lambda_{\langle{}ij\rangle}$ between neighboring sites for the central columns in a AXC4 cylinder of length $L_x=24$. 
			The red dashed lines represent the 1D path in DMRG calculations, and plaquette SU(4)-singlet projectors are defined on the rhombuses. 
			(b) The two-plaquette SU(4)-singlet correlation function $D(r)$ as a function of distance on AXC4 cylinders. Two plaquettes $P_{p}$ and $P_{p+r\hat{x}}$ are set on the same row.}
	\end{figure}

	\subsection{$\Lambda_{\langle{}ij\rangle}$ on odd circumference cylinders}
	We further remark that the behaviors of the ground state of the SU(4) KK model on odd circumference cylinders are very sensitive to the type and width of the cylinder. 
	For $L_{y} = 3$ cylinders, previous study in XC3 geometry found that the nearest-neighbor correlation $\Lambda_{\langle{}ij\rangle}$ is quite uniform~\cite{Xu2020}.
	However, for the YC3 geometry, our results indicate that the pattern of $\Lambda_{\langle{}ij\rangle}$ still spatially fluctuates despite of good convergence in DMRG simulations [see Fig.~\ref{fig:sm-odd-cyl-bondeng}(a)]. 
	For a wider YC5 cylinder, $\Lambda_{\langle{}ij\rangle}$ is also strongly fluctuating and has fourfold oscillations [see Fig.~\ref{fig:sm-odd-cyl-bondeng}(b)]. 
	Although this fluctuation is suppressed on the XC5 cylinder, a small inhomogeneity of $\Lambda_{\langle{}ij\rangle}$ is clearly visible [see Fig.~\ref{fig:sm-odd-cyl-bondeng}(c)].
	\begin{figure}[ht]
		\centering
		\includegraphics[width=0.72\linewidth]{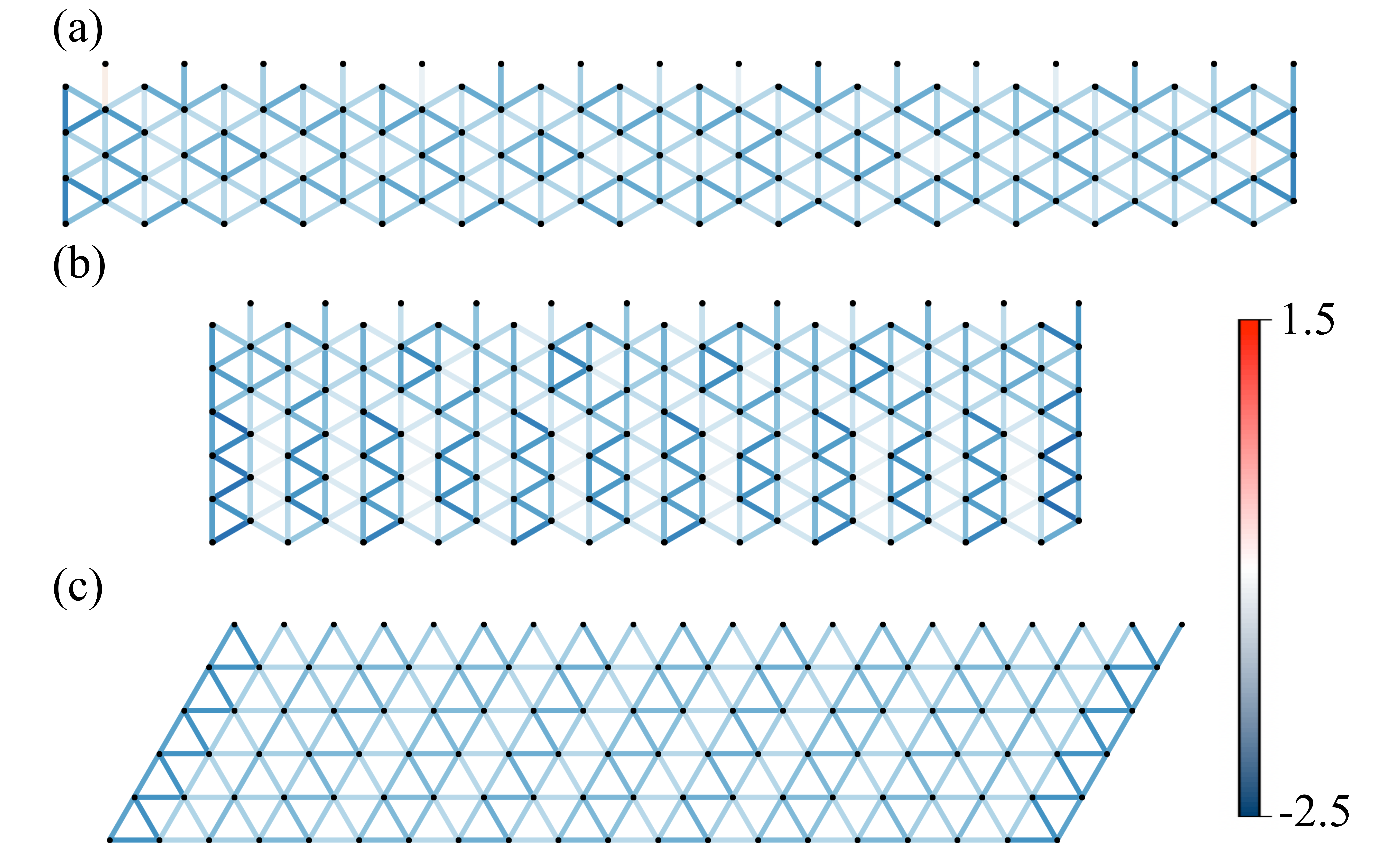}
		\caption{\label{fig:sm-odd-cyl-bondeng} The expectation value $\Lambda_{\langle{}ij\rangle}$ between neighboring sites in odd circumference cylinders: (a) YC3 with $L_{x} = 16$, (b) YC5 with $L_{x} = 12$, and (c) XC5 with $L_{x} = 20$. The bond dimension used in DMRG calculations is set up to $18000 \sim 20000$.}
	\end{figure}

	\section{Projected uniform $\pi$-flux state on cylinder and torus}
	Although the Gutzwiller projected uniform $\pi$-flux state exhibits ``translational symmetry breaking'' on cylinders in viewing from $\Lambda_{\langle{}ij\rangle}$, the twofold oscillation is smeared out in the central region of the cylinder [see Fig.~\ref{fig:sm-uniformpi}(a)]. 
	This smearing also appears in the oscillation of the bipartite EE $S(l_{x})$, as shown in Fig.~\ref{fig:sm-uniformpi}(c).
	Notice that the bipartite $S(l_x)$ of the projected uniform $\pi$-flux state does not saturate, which is sharply different from those shown in Fig.~\eqref{fig:sm-eex}.
	We further place the projected uniform $\pi$-flux state on a $L_x\times{}L_y=6\times{}4$ torus and measure the nearest-neighbor correlation $\Lambda_{\langle{}ij\rangle}$. 
	The uniform pattern of $\Lambda_{\langle{}ij\rangle}$ illustrated in Fig.~\ref{fig:sm-uniformpi}(b) indicates that the translational symmetry is not broken. Thus, the ``translational symmetry breaking'' observed on cylinders is more likely an open-boundary effect.
	\begin{figure}[ht]
		\centering
		\includegraphics[width=0.8\linewidth]{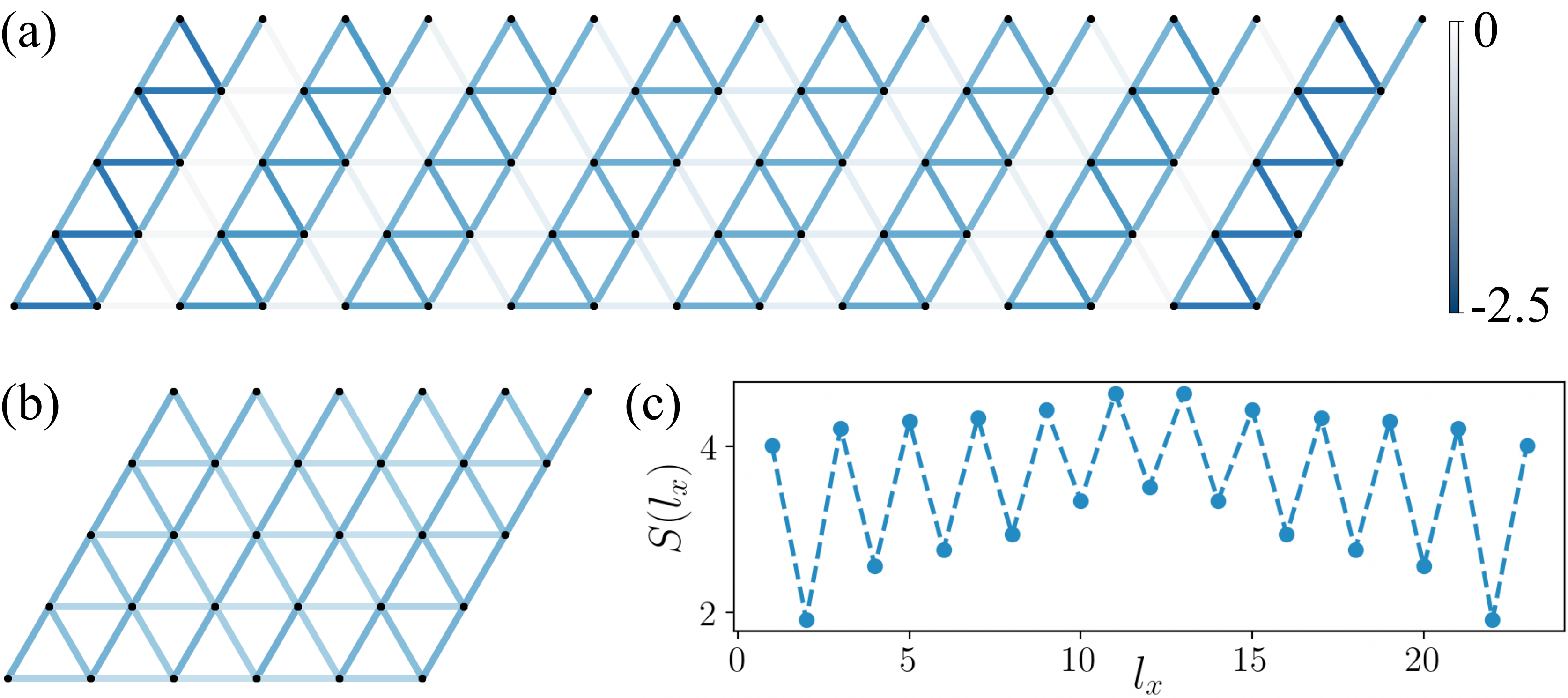}
		\caption{\label{fig:sm-uniformpi} The nearest-neighbor correlation $\Lambda_{\langle{}ij\rangle}$ for the projected uniform $\pi$-flux state on XC4 (a) cylinder with $L_x = 16$ and (b) torus with $L_x = 6$. (c) The bipartite EE $S(l_{x})$ versus subsystem length $l_{x}$ for the projected uniform $\pi$-flux state on an XC4 cylinder with length $L_x=24$.}
	\end{figure}
\end{widetext}

\end{document}